\documentclass[
]{ceurart}

\usepackage{listings}
\usepackage{svg}
\usepackage{tabularx}
\begin{document}

\copyrightyear{2025}
\copyrightclause{Copyright for this paper by its authors.
  Use permitted under Creative Commons License Attribution 4.0
  International (CC BY 4.0).}

\conference{CLEF 2025 Working Notes, 9 -- 12 September 2025, Madrid, Spain}

\title{DS@GT at CheckThat! 2025: Exploring Retrieval and Reranking Pipelines for Scientific Claim Source Retrieval on Social Media Discourse}

\title[mode=sub]{Notebook for the CheckThat! Lab at CLEF 2025}

\author[1]{Jeanette Schofield}[
orcid=0009-0000-0669-8962,
email=jschofield8@gatech.edu
]
\author[1]{Shuyu Tian}[
orcid = 0000-0001-6444-651X,
email=stian40@gatech.edu
]
\author[1]{Hoang Thanh Thanh Truong}[
orcid=0009-0007-4130-3349,
email=htruong47@gatech.edu
]
\author[1]{Maximilian Heil}[
orcid=0009-0002-6459-6459,
email=mheil7@gatech.edu
]
\cormark[1]

\address[1]{Georgia Institute of Technology, North Ave NW, Atlanta, GA 30332}
\cortext[1]{Corresponding author.}

\begin{abstract}
    Social media users often make scientific claims without citing where these claims come from, generating a need to verify these claims. This paper details work done by the DS@GT team for CLEF 2025 CheckThat! Lab Task 4b Scientific Claim Source Retrieval which seeks to find relevant scientific papers based on implicit references in tweets. Our team explored 6 different data augmentation techniques, 7 different retrieval and reranking pipelines, and finetuned a bi-encoder. Achieving an MRR@5 of 0.58, our team ranked 16th out of 30 teams for the CLEF 2025 CheckThat! Lab Task 4b, and improvement of 0.15 over the BM25 baseline of 0.43. Our code is available on Github at \url{https://github.com/dsgt-arc/checkthat-2025-swd/tree/main/subtask-4b}.
\end{abstract}


\maketitle

\section{Introduction}

The spread of health misinformation on social media has increasingly become a problem in recent years creating the need to substantiate claims made by users on social media \cite{kbaier2024health}. CheckThat! Task 4b, Scientific Claim Source Retrieval, attempts to solve this problem by asking users to retrieve the most relevant scientific articles from a collection set that support user claims on social media. This is a hard challenge to solve as the language used on social media often differs greatly from that used in the scientific articles that make up the collect set \cite{hafid:citeworthiness}.

The DS@GT team explored finetuning a bi-encoder, 6 different data augmentation techniques, and 7 different retrieval and reranking pipelines. Achieving an MRR@5 of 0.58, our team ranked 16th out of 30 teams for the CLEF 2025 CheckThat! Lab Task 4b, an improvement of 0.15 over the BM25 baseline of 0.43.

\section{Related Work}
First offered in 2018, CheckThat! 2025 is the eighth offering of the CheckThat! Lab which focuses on technology that helps automate the journalistic verification process \cite{clef-checkthat:2025:task4, clef2025-workingnotes}. Task 4b, Scientific Claim Source Retrieval, predicts which scientific articles are most relevant to user discourse related to COVID-19 on Twitter. Prior CheckThat! labs have offered COVID-related tasks. The goal of Task 1, Identifying Relevant Claims in Tweets, in 2022 was to predict which COVID-19 related tweets in a dataset were worth fact-checking \cite{clef-checkthat:2022:task1, clef2022-workingnotes}. Task 4, Detecting hero, villain, and victim from memes, from CheckThat! 2024 asked participates to identify the role of entities within memes, many of which were related to COVID-19 \cite{clef2024-workingnotes}. 

Before a source can be retrieved to support a claim, one first needs to determine if a claim is being made. Cite-worthiness refers to the problem of determining if there are missing references to scientific results in text. Designed for cite-worthiness detection in scientific text, the CiteWorth dataset \cite{wright:citeworth} contains 1.1M English-language sentences, with 375K sentences designed as cite-worthy. Models trained using the CiteWorth dataset were found to perform poorly when tasks with scientific citations in social media discourse \cite{hafid:citeworthiness}. Consisting of 1,261 tweets and a subset of the SciTweets dataset, SCiteTweets is dataset created for detecting citations in social media discourse \cite{haid:scitweets}.

\section{Data}
\begin{figure}[htbp]
    \centering
    \begin{minipage}{0.48\linewidth}
        \centering
        \includegraphics[width=\linewidth]{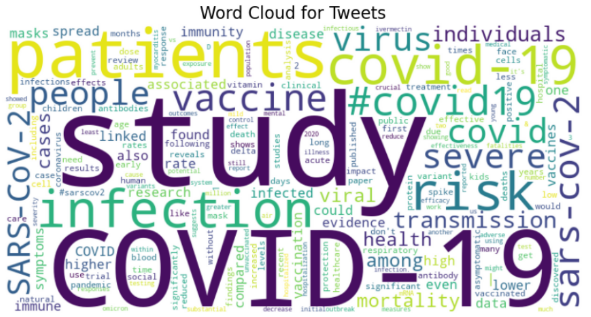}
        \caption{Word Cloud of Tweets in dataset}
        \label{fig:tweets-wordcloud}
    \end{minipage}
    \hfill
    \begin{minipage}{0.48\linewidth}
        \centering
        \includegraphics[width=\linewidth]{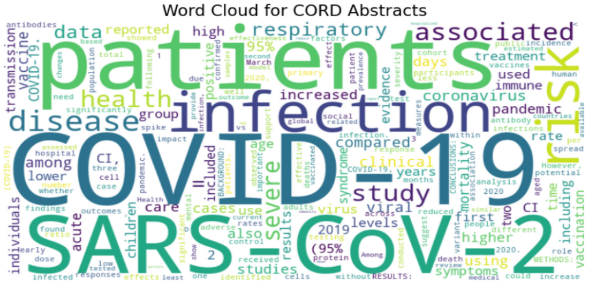}
        \caption{Word Cloud of Abstracts in dataset}
        \label{fig:abstracts-wordcloud}
    \end{minipage}
\end{figure}

Task 4b organizers provided three datasets: two sets of query tweets, one for training and one for development, and the CORD-19 paper collection. The training (train) dataset contains 12,853 tweets, while the development (dev) set includes 1,400 tweets. Both datasets share the same data structure with the following three columns:
\begin{enumerate}
    \item post\_id - A unique identifier for the tweet
    \item tweet\_text - The textual content of the tweet
    \item cord\_uid - A unique identifier corresponding to a paper in the CORD-19 collection
\end{enumerate}
The CORD-19 paper collection consists of 7,718 academic papers with 17 columns. Some of the key columns are:
\begin{enumerate}
    \item cord\_uid - A unique identifier for the paper for linking with tweet queries in the train and dev set
    \item title - The title of the paper
    \item abstract - The abstract of the paper
    \item authors - A list of the paper's authors
    \item journal - The journal where the paper was published
    \item publish\_time - The publication date
\end{enumerate}
Word clouds in Figures \ref{fig:tweets-wordcloud} and \ref{fig:abstracts-wordcloud} show that words such as "patients", "COVID-19", "infection", and "risk" frequently appear in the set of query tweets and the CORD-19 paper collection. Words such as "study" and "vaccine" appear often in the tweets dataset, but are not as prevalent in paper abstracts. Terms like "disease" and "pandemic" appear more often in the CORD-19 paper abstracts than in the query tweets.

\section{Methodology}
For this task, we explored multiple retrieval and reranking pipelines, experimenting with BM25 and bi-encoders for retrieval and exploring 7 different models for reranking. Multiple data augmentation experiments were performed during the retrieval stage to see if modifying the tweet data would improve results.

\subsection{Mean Reciprocal Rank (MRR)}
The Mean Reciprocal Rank at k (MRR@k) was the official metric for the CheckThat! 2025 Task 4b \cite{clef-checkthat:2025:task4}. For a set of N queries, the MRR is calculated as:

\begin{equation}
    MRR@k=\frac{1}{N}\sum^{N}_{i=1}\frac{1}{rank_i}
\end{equation}

where \(rank_i\) is the position of the first correct and relevant document in the top-k retrieved list for the i-th query. If the relevant document does not appear within the top-k results, the reciprocal rank for that query is treated as zero.

\subsection{Retrieval and Reranking Pipeline: First Attempt}

Our first attempt at a retrieval and reranking pipeline used BM25 for retrieval, building on the organizers' baseline. BM25, or Best Match 25, ranks documents based on how many query terms appear in each document \cite{wikipediaOkapiBM25}. BM25 is often used to return a subset of documents from a large corpus due to it's speed. After the initial subset of documents are retrieved, more computationally expensive reranking algorithms are applied to improve upon the initial results. In our first pipeline, a cross-encoder trained on the MS Marco dataset (cross-encoder/ms-marco-MiniLM-L6-v2) was used for reranking \cite{nguyen2016msmarco, cross_encoder_ms_marco_MiniLM_L6_v2}.

In all pipelines created, 100 documents were retrieved. Prior to retrieval, tweets and documents were converted to lowercase, tokenization was applied by splitting sentences up into words (i.e. on spaces), and stop words were removed. Data from the tweet\_text column was used for the query set. For the CORD-19 paper collecition, we combined the title of the paper with the abstract.

\subsection{Exploring Data Augmentation}

After setting up an initial retrieval and reranking pipeline, data augmentation techniques were explored. For Task 4b, the goal was to find the journal article that best matched a user tweet from Twitter. Language used in social media discourse tends to be very different than that used in academic writing. Social media discourse often takes a more informal tone whereas academic writing uses formal language.

With this in mind, we explored various means of augmenting the tweet\_text in the query dataset: rewriting tweets using formal (i.e. scientific) language, concatenating the original tweet text with the rewritten tweets, and replacing the original tweet with science-related keywords used in the tweet.

Two experiments replaced the original tweet text with rewritten text. The first experiment, "Replace w/ Formal Rewritten", used the prompt "Rewrite the input using formal language" to generate the rewritten tweets. Upon noticing that some of the tweets in the query set were not written in English, a second experiment was performed. "Replace w/ English Formal Rewritten" used the prompt "Rewrite the input using formal English language."

Three data augmentation experiments involved combining the original tweet text with the rewritten text. "Concat w/ Formal" combined the original tweet with tweets were rewritten using "formal language" prompt. "Concat w/ English formal" combined the original tweet with tweets rewritten using the "formal English language" prompt. "Concat w/ All" combined the original tweet, the tweet rewritten using "formal language", and the tweet rewritten using "formal English language."

The experiment "Replace w/ Keywords" used the prompt "Return a list of only science-related keywords in the tweet." The motivation was to see whether or not our pipeline would improve by using only science-related keywords to retrieve relevant documents.

Table \ref{table:examples_data_augmentation} provides an example of the augmented data for each experiment.

\begin{table}[h]
\centering
\caption{Data Augmentation Experiments and Generated Tweet Text. \\
Data generated from the tweet with a post\_id of 3491 in the query set}
\label{table:examples_data_augmentation}
\begin{tabularx}{\textwidth}{|>{\hsize=0.25\hsize}X|>{\hsize=0.75\hsize}X|}
\hline
\textbf{Data Augmentation Experiment} & \textbf{Tweet Text} \\
\hline
Original Dataset & Bile salts in gut and liver pathophysiology \\
Replace w/ Formal Rewritten & Bile salts in the pathophysiology of the gastrointestinal tract and hepatic systems. \\
Replace w/ English Formal Rewritten & The role of bile salts in the pathophysiology of the gastrointestinal tract and liver. \\
Concat w/ Formal & Bile salts in gut and liver pathophysiology Bile salts in the pathophysiology of the gastrointestinal tract and hepatic systems. \\
Concat w/ English Formal & Bile salts in gut and liver pathophysiology The role of bile salts in the pathophysiology of the gastrointestinal tract and liver. \\
Concat w/ All (Formal \& English Formal) & Bile salts in gut and liver pathophysiology The role of bile salts in the pathophysiology of the gastrointestinal tract and liver. Bile salts in the pathophysiology of the gastrointestinal tract and hepatic systems. \\
Replace w/ Keywords & Bile, salts, gut, liver, pathophysiology \\
\hline
\end{tabularx}
\end{table}

In all data augmentation experiments, BM25 was used for retrieval and cross-encoder/ms-marco-MiniLM-L6-v2 was used for reranking. gpt-4o was used in all experiments to rewrite tweet data.

\subsection{Experimenting with Reranking Models}

After our data augmentation techniques, we changed the retrieval stage to use BM25-PyTorch instead of BM25. This was done to make use of GPU and run the pipeline faster. With BM25-PyTorch as our retrieval model, we explored 7 different reranking models to see how they performed.

Three of these reranking models were trained on the MS Marco dataset: cross-encoder/ms-marco-MiniLM-L-6-v2, tomaarsen/reranker-msmarco-ModernBERT-base-lambdaloss, and tomaarsen/reranker-msmarco-MiniLM-L12-H384-uncased-lambdaloss \cite{nguyen2016msmarco, cross_encoder_ms_marco_MiniLM_L6_v2, tomaarsen_reranker_msmarco_modernbert_base_lambdaloss, tomaarsen_reranker_msmarco_minilm_l12_h384_uncased_lambdaloss}. MS Marco is a widely used question answer dataset featuring 100,000 Bing questions with huamn generated answers.

In an effort to explore models trained on different datasets, we tested our pipeline on 3 models trained on the GooAQ dataset which contains 5 million Google queries and 3 million answers: akr2002/reranker-ModernBERT-base-gooaq-bce, tomaarsen/reranker-ModernBERT-large-gooaq-bce, and tomaarsen/reranker-NeoBERT-gooaq-bce \cite{khashabi2021gooaq, akr2002_reranker_modernbert_base_gooaq_bce, tomaarsen_reranker_modernbert_large_gooaq_bce, tomaarsen_reranker_neobert_gooaq_bce}.

Our last experiment for exploring retrieval and reranking pipelines used Google's T5 (Text-to-Text Transfer Transformer) model which is trained on the C4 dataset which was developed by Google and Meta via scraping the web \cite{raffel2020exploring, aiaaic_c4_dataset}. We used AnswerDotAI's Reranker library to implement this model \cite{clavie2024rerankers_repo, clavie2024rerankers_paper}.

\subsection{Using Bi-encoders for Retrieval}

\begin{figure}[ht]
    \centering
    \includegraphics[width=1.0\linewidth]{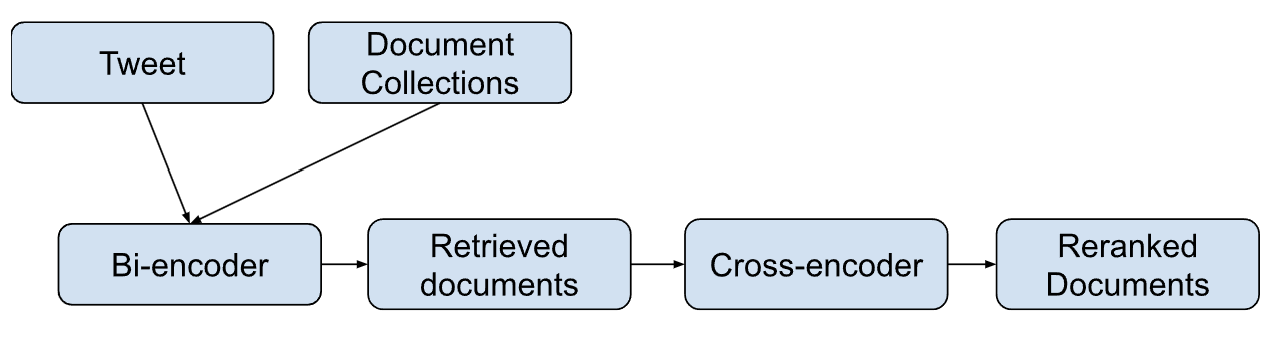}
    \caption{Two-stage document retrieval pipeline.}
    \label{fig:bi_cross_encoder}
\end{figure}

In this section, we focus on the bi-encoder and cross-encoder approaches using bi-encoder/msmarco-distilbert-base-v4 and cross-encoder/ms-marco-MiniLM-L-6-v2, respectively \cite{sentence_transformers_msmarco_distilbert_base_v4, cross_encoder_ms_marco_MiniLM_L6_v2}. Both are provided by the Sentence-Transformers framework \cite{reimers-2019-sentence-bert}. We aim to leverage the speed of bi-encoders to narrow down a smaller pool of document candidates and then use the precision of cross-encoders to enhance document retrieval accuracy (see Figure~\ref{fig:bi_cross_encoder}).

In the first stage, the pipeline loads a data set of query tweets and a corpus of scientific papers comprising titles and abstracts. These inputs are processed by the bi-encoder model (msmarco-distilbert-base-v4) to generate dense vector embeddings. Then, we perform a semantic search to retrieve the top 100 most similar documents for each tweet based on cosine similarity. 

Since the cross-encoder is computationally expensive, this initial retrieval step helps reduce the relevant scientific documents to save resources. In the second stage, these top 100 documents are reranked using the cross-encoder (ms-marco-MiniLM-L-6-v2) to improve retrieval accuracy.

We evaluate both stages using MRR@k to allow a direct comparison between the initial retrieval and reranking performance.

\subsection{Bi-encoder Finetuning}

Building on the strong performance of the bi-encoder and cross-encoder pipeline, we finetuned the bi-encoder to further improve retrieval quality. 

To expand the training data, we sample sentence-pairs to generate a silver dataset using contextual text augmentation, following the strategy proposed in Augmented SBERT \cite{thakur-2020-AugSBERT}. We load tweets paired with its corresponding scientific document, which includes the document's title and abstract. 

Next, we process the input data using a BERT-based augmenter (ContextualWordEmbsAug with bert-base-uncased) to insert contextually appropriate words. These augmented tweet–document pairs are saved as silver examples to diversify data and improve the generalization of the model.

After augmenting data, we train our bi-encoder (msmarco-distilbert-base-v4) through two phases. We firstly train only on the train data set and then train on the silver dataset, measured via MRR@100 on the dev set.

After data augmentation, we finetune our bi-encoder model (msmarco-distilbert-base-v4) in two phases. In the first phase, we train the model exclusively on the original training dataset. In the second phase, we continue training on the silver dataset generated through augmentation. 

Both models are trained and evaluated using the same procedure. We employ the SentenceTransformerTrainer interface with a contrastive cosine similarity loss, and evaluate model performance using Mean Reciprocal Rank at 100 (MRR@100) on the dev set. The best models with highest MRR@100 are saved at each phase.

Together, this two-phase training strategy enables the model to first learn from gold-labeled data and then improve generalization through silver training examples generated via contextual augmentation. This progressive finetuning pipeline is designed to enhance retrieval performance while maintaining robustness on unseen tweet–document pairs.

\section{Results}

For the final submission for Task 4b, our team used BM25-Pytorch for retrieval and T5 for reranking. No data augmentation was applied. This pipeline achieved an MRR@5 of 0.58 which improved over the baseline of 0.43 by 0.15. Our team ranked 16 out of 20 teams. The leaders, SourceSniffers, achieved an MRR@5 of 0.68.

\subsection{Data Augmentation}

As shown in table \ref{table:results_data_augmentation}, data augmentation experiments that replaced the text of the original tweet performed worse than the baseline on reranking tasks, with the MRR@5 decreasing by 0.06 or more. Experiments that combined the original tweet with additional data showed a slight improvement over the baseline with an increase in MRR@5 of roughly 0.03 after retrieval. After reranking, the experiment "Concat w/ Formal" and "Concat w/ All" improved over the baseline by 0.01.

\begin{table}[h]
\centering
\caption{Data Augmentation Results \\
BM25 was used for retrieval and the cross-encoder ms-marco-MiniLM-L-6-v2 was used for reranking}
\label{table:results_data_augmentation}
\begin{tabularx}{\textwidth}{|>{\hsize=0.435\hsize}X|>{\hsize=0.29\hsize}X|>{\hsize=0.275\hsize}X|}
\hline
\textbf{Data Augmentation Step} & \textbf{MRR@5 after Reranking}  & \textbf{MRR@5 after Retrieval} \\
\hline
None & 0.5521 & 0.6028 \\
Replace w/ Formal Rewritten & 0.4915 & 0.5183 \\
Replace w/ English Formal Rewritten & 0.5112 & 0.5366 \\
Concat w/ Formal & 0.5823 & 0.6106 \\
Concat w/ English Formal & 0.5859 & 0.6092 \\
Concat w/ All (Formal \& English Formal) & 0.5812 & 0.5618 \\
Replace w/ Keywords & 0.4280 & - \\
\hline
\end{tabularx}
\end{table}

\subsection{Reranking Models}

Table \ref{table:results_reranking_models} shows the results for experimenting with various reranking models. In all experiments, BM25-Pytorch was used the as the model for retrieval. Of the 7 models used for reranking, only the cross-encoder ms-marco-MiniLM-L-6-v2 model and the T5 model improved upon the initial retrieval results. The 3 models trained on the MS Marco dataset performed significantly better than those trained on the GooAQ dataset.

\begin{table}[h]
\centering
\caption{Experimenting with Reranking Models Results\\
BM25-Pytorch was used for retrieval on all experiments. \\
*Submitted to CLEF for Task 4b
}
\label{table:results_reranking_models}
\begin{tabularx}{\textwidth}{|>{\hsize=0.6\hsize}X|>{\hsize=0.2\hsize}X|>{\hsize=0.2\hsize}X|}
\hline
\textbf{Reranking Model} & \textbf{MRR@5 After Retrieval} & \textbf{MRR@5 After Reranking} \\
\hline
ms-marco-MiniLM-L-6-v2 & 0.6300 & 0.6474 \\
reranker-msmarco-ModernBERT-base-lambdaloss & 0.6300 & 0.6219 \\
reranker-msmarco-MiniLM-L12-H384-uncased-lambdaloss & 0.6300 & 0.4194 \\
reranker-ModernBERT-base-gooaq-bce & 0.6300 & 0.2437 \\
reranker-ModernBERT-large-gooaq-bce & 0.6300 & 0.5471 \\
reranker-NeoBERT-gooaq-bce & 0.6300 & 0.0563 \\
T5* & 0.6300 & 0.6590 \\
\hline
\end{tabularx}
\end{table}

\subsection{Bi-encoder Finetuning}

Results for bi-encoder finetuning can be found in table \ref{table:results_finetuning}. Without finetuning, the MRR@5 using a bi-encoder for retrieval was 0.428. Unfortunately, all 3 finetuning experiments performed worse than the bi-encoder without any finetuning. The gold dataset with hard negatives from silver data achieved an MRR@5 of 0.407 which is 0.011 less than the regular bi-encoder.

\begin{table}[h]
\centering
\caption{Finetuning Bi-encoder Results\\
Cross-encoder model ms-marco-MiniLM-L-6-v2 was used for reranking on all experiments
}
\label{table:results_finetuning}
\begin{tabularx}{\textwidth}{|>{\hsize=0.6\hsize}X|>{\hsize=0.2\hsize}X|>{\hsize=0.2\hsize}X|}
\hline
\textbf{Finetuned On} & \textbf{MRR@5 After Retrieval} & \textbf{MRR@5 After Reranking} \\
\hline
No Finetuning & 0.428 & 0.612 \\
Gold dataset & 0.178 & 0.445 \\
Gold dataset + silver augmented data & 0.343 & 0.568 \\
Gold dataset + hard negatives from silver data & 0.407 & 0.616 \\
\hline
\end{tabularx}
\end{table}

\section{Discussion}

\subsection{Data Augmentation}
For data augmentation, all 3 experiments that replaced the original tweet text with rewritten text caused the MRR@5 to decrease. This suggests that the original tweets contained context that was lost by rewriting the tweet. This appears to be especially true for the "Replace w/ Keywords" experiment that removed all words from the tweet that were not related to science in some way.

The increase in the MRR@5 after reranking and retrieval for experiments where the original tweet was combined with the rewritten tweet suggests that multiple representations of the same tweet may improve performance.

Finally, it is worth noting that our data augmentation experiments were only applied to the query dataset. The data for the CORD-19 paper collection as not manipulated in any way. For some experiments, this probably affected results. For example, the "Replace w/ Keywords" likely would have performed better if the prompt "Return a list of only science-related keywords in the tweet" was applied to both the query and CORD-19 paper collection.

\subsection{Reranking Models}
As noted in the results section, 3 the models trained on the MS Marco dataset performed significantly better than those trained on the GooAQ dataset. The MS Marco dataset was generated from Bing queries with human-generated answers. The GooAQ dataset was curated using queries from Google with answers drawn from Google's responses to questions. Both datasets are generated from search queries, but the way the answers were generated differs greatly. One possible explanation is that the human-generated answers to the MS Marco dataset more closely reflect the language used in the abstracts for the CORD-19 paper collection. Another explanation might stem from differences in how users ask questions on Bing and Google. Perhaps users on Bing tend to be more verbose or search for different criteria than Google users.

\subsection{Bi-encoder Finetuning}
finetuning the bi-encoder decreased retrieval performance on all experiments. However, the "Gold dataset + hard negatives from silver data" was very close to the baseline, suggesting that there is the potential for improvement with further exploration.

\section{Future Work}
Task 4b leaders SourceSniffers achieved an MRR@5 of 0.68. The problem of finding relevant scientific claims for social media discourse is still unsolved. Future work should prioritize improving upon the work CheckThat! Task 4b participants explored this year.

\subsection{Data Augmentation}
As noted earlier, data augmentation techniques were only applied to query tweets. To see if data augmentation yields helpful results, these techniques should be applied to both the query dataset and the CORD-19 paper collection.

For the CORD-19 paper collection, it would be interesting to see if summarizing abstracts affects model performance. The abstracts in the CORD-19 paper collection were anywhere from 2 words to 1800 words. If social media users are more likely to reference mains ideas from the papers in their tweets, summarizing abstracts might lead to improved retrieval results.

In our experiments, only the title and abstract from the CORD-19 paper collection were used to when looking for relevant articles. It would be interesting to see if including additional features like the journal that the article appeared in and the names of authors would improve performance.

\subsection{Bi-encoder Finetuning}

The bi-encoder performed worse after finetuning, suggesting that the silver dataset may introduce noise that hinders the model's ability to retrieve relevant documents. Future work should focus on filtering the silver data before finetuning. We could explore a pre-trained cross-encoder to score the augmented tweet–document pairs and retain only high-confidence examples. The results highlight that more data may not lead to better performance, especially when additional examples are not meaningfully contributed to the learning signal.

\subsection{Other Avenues to Explore}
If retrieval results were poor, reranking could only do so much. Future work should prioritize techniques that improve retrieval. One potential avenue for this is to combine multiple approaches: augmenting data, finetuning a bi-encoder for retrieval, and finding the best model for reranking.

\section{Conclusion}
This paper demonstrates the work performed by DS@GT for CheckThat! 2025 Task 4b, Scientific Claim Source Retrieval. We explored data augmentation techniques, reranking and retrieval pipelines, and finetuning bi-encoders. We found that combining original tweets with rewritten tweets that reflect the language of the document collection may lead to an improvement in retrieval. Using BM25-Pytorch for retrieval and T5 for reranking, we achieved an MRR@5 of 0.58 on the evaluation set, and improvement of 0.15 over the baseline of 0.43. Future work should prioritize improving the retrieval stage of the retrieval and reranking pipeline. Our code is available on Github at \url{https://github.com/dsgt-arc/checkthat-2025-swd/tree/main/subtask-4b}

\section*{Acknowledgments}

Thank you to the everyone in the DS@GT CLEF team for their support. Special thanks to Anthony Miyaguchi and Murilo Gustenelli for their many hours of work organizing and supporting the DS@GT CLEF team. This paper would not have happened without you!

Thank you to Partnership for an Advanced Computing Environment (PACE) \cite{PACE} at the Georgia Institute of Technology, Atlanta, Georgia, USA for allowing us to use their resources to perform this research.

\section*{Declaration on Generative AI}
Generative AI was used in the preparation of this work, specifically to format bibliography references and for formatting tables and figures. After using these tools, the authors reviewed and edited the content as needed and take full responsibility for the publication’s content.

\bibliography{sample-ceur}

\appendix

\end{document}